\documentclass[conference]{IEEEtran}
\usepackage{booktabs}
\IEEEoverridecommandlockouts
\usepackage{cite}
\usepackage{amsmath,amssymb,amsfonts}
\usepackage{algorithmic}
\usepackage[utf8]{inputenc} 
\usepackage{graphicx}
\usepackage{url}
\usepackage{textcomp}
\usepackage{xcolor}
\def\BibTeX{{\rm B\kern-.05em{\sc i\kern-.025em b}\kern-.08em
T\kern-.1667em\lower.7ex\hbox{E}\kern-.125emX}}

\newcommand{\edit}[1]{\textcolor{black}{#1}}

\title{Investigating the Use of LLMs for Evidence Briefings Generation in Software Engineering\\}

\author{
\IEEEauthorblockN{
\begin{tabular}{ccc}
Mauro Marcelino & Marcos Alves & Bianca Trinkenreich \\
\textit{PUC-Rio} & \textit{FITec Technological Innovations} & \textit{Colorado State University} \\
Rio de Janeiro, Brazil & Belo Horizonte, Brazil & Fort Collins, USA \\
moliveira@inf.puc-rio.br & marcosaalves@fitec.org.br & Bianca.trinkenreich@colostate.edu
\end{tabular}
\vspace{1em}
\\
\begin{tabular}{cccc}
Bruno Cartaxo & Sérgio Soares & Simone D.J. Barbosa & Marcos Kalinowski \\
\textit{IFPE} & \textit{UFPE} & \textit{PUC-Rio} & \textit{PUC-Rio} \\
Recife, Brazil & Recife, Brazil & Rio de Janeiro, Brazil & Rio de Janeiro, Brazil \\
email@brunocartaxo.com & scbs@cin.ufpe.br & simone@inf.puc-rio.br & kalinowski@inf.puc-rio.br
\end{tabular}
}
}

\begin{document}

\maketitle

\begin{abstract}

[Context] An evidence briefing is a concise and objective transfer medium that can present the main findings of a study to software engineers in the industry. Although practitioners and researchers have deemed Evidence Briefings useful, their production requires manual labor, which may be a significant challenge to their broad adoption. [Goal] The goal of this registered report is to describe an experimental protocol for evaluating LLM-generated evidence briefings for secondary studies in terms of content fidelity, ease of understanding, and usefulness, as perceived by researchers and practitioners, compared to human-made briefings. [Method] We developed a RAG-based LLM tool to generate evidence briefings. We used the tool to automatically generate two evidence briefings that had been manually generated in previous research efforts. We designed a controlled experiment to evaluate how the LLM-generated briefings compare to the human-made ones regarding perceived content fidelity, ease of understanding, and usefulness. [Results] To be reported after the experimental trials. [Conclusion] Depending on the experiment results.

\end{abstract}


\section{Introduction}

Evidence-Based Software Engineering (EBSE), introduced by Kitchenham \textit{et al.}~\cite{Kitchenham2004}, is an adaptation of evidence-based principles from medicine designed to bring scientific rigor to the practice of Software Engineering (SE). It emphasizes the systematic collection, evaluation, and synthesis of empirical evidence to support decision-making processes, mainly in the form of Systematic Literature Reviews (SLR). The core objectives of SLRs include summarizing existing evidence, identifying research gaps, and establishing an empirical foundation to inform both academic inquiry and industry practices.

However, one of the main challenges in SE research is the limited applicability in industrial contexts \cite{winters2024}. Studies have shown that research findings are often misaligned with industry needs and do not adequately reflect practical constraints \cite{garousi2020}. With respect to SLRs, Santos and Silva~\cite{santos2013motivation} reported that only a small fraction of SE researchers see their SLR directly influencing industry practices. Only six out of the 44 researchers interviewed claimed that their work had somehow influenced industry practice. \edit{Similarly, Budgen \textit{et al.}~\cite{budgen2020support} examined 49 studies and found that while SLRs aggregate valuable insights, their translation into actionable knowledge remains a challenge, indicating that the limited industrial impact represents a persistent trend}. 
Other works have highlighted ineffective mechanisms for transferring knowledge into actionable insights as a key factor contributing to this gap \cite{akdur2019design,garousi2019aligning}.

These findings underscore the misalignment between research priorities and real-world industry needs, 
highlighting the need for better mechanisms to translate academic research into practical applications. Ensuring that research findings are reliable, easily accessible, and actionable for industry professionals remains a critical challenge. Beecham \textit{et al.}~\cite{Beecham2014}, for instance, emphasized this issue by arguing that the medium used to communicate research findings to practitioners plays an important role in bridging this gap. 
To address these limitations, Cartaxo \textit{et al.}~\cite{Cartaxo2016} introduced evidence briefings as an alternative knowledge transfer mechanism in SE. 
Evidence briefings condense the most relevant findings from empirical studies into concise, structured, and accessible reports aimed at industry practitioners. However, despite their potential benefits, the adoption of evidence briefings remains limited. The manual effort and cognitive workload involved in their creation lead to scalability issues, hindering widespread adoption in both research and industrial settings.

The advancement of artificial intelligence has introduced new possibilities for automating research synthesis. Machine Learning (ML) algorithms have historically been used in text summarization, but Large Language Models (LLM) \cite{attention} have taken this a step further by leveraging deep contextual understanding and pre-trained knowledge to extract and synthesize complex text~\cite{zhang2024}. Recent studies have demonstrated the potential of LLMs for automating the creation of research synthesis in different domains, including summarizing medical evidence \cite{tang2023evaluating} and the generation of policy-oriented summaries based on research~\cite{rosenfeld2024building}.


While these aforementioned studies indicate that LLMs have the potential to reduce manual effort and improve accessibility in research synthesis, challenges remain. LLMs are known to generate hallucinated content, \textit{i.e.} where the model generates information that is factually incorrect or not present in the source text \cite{zhang2024}, \edit{require careful prompt engineering} \cite{geroimenko2025key}, and may sometimes produce overly confident or misleading summaries \cite{tang2023evaluating}. These issues raise concerns regarding the reliability and applicability of AI-generated research summaries, particularly in high-stakes decision-making. 

Additionally, existing automated evaluation metrics, such as ROUGE \cite{rouge} and BertScore \cite{bertscore}, have been shown to correlate weakly with human assessments in key aspects like coherence and factual consistency \cite{fabbri2021summevalreevaluatingsummarizationevaluation}. Similarly, Tang \textit{et al.} \cite{tang2023evaluating} found that existing automatic metrics fall short in accurately measuring factual inconsistency and human preferences in medical evidence summarization, \edit{further reinforcing the need for empirical studies involving humans}. 

In this study, we investigate whether automatically generated Evidence Briefings can be suitable for broader adoption in SE. Therefore, we designed an experiment \edit{with human subjects} following the recommendations by Wohlin \textit{et al.}~\cite{wohlin2024experimentation}.
The main goal of the experiment is to assess whether LLMs can generate Evidence Briefings that are faithful to the evidence contained in the original text and easily digestible by practitioners.

This registered report is organized as follows: Section \ref{sec:background} discusses Evidence Briefings in SE and the use of LLMs for research synthesis; Section \ref{sec:experiments} describes the experimental protocol in detail; and Section \ref{sec:nextsteps} presents the next steps to conclude the study.

\section{Background and Related Work}
\label{sec:background}

\subsection{Evidence Briefings in SE}

An evidence briefing is a one-page document that summarizes the key findings of an empirical study \cite{Cartaxo2018}. Its concise format is essential for presenting evidence in a clear and engaging manner, making it more accessible to both practitioners and researchers. Compared to SLR papers, evidence briefings provide a more practical and efficient medium for acquiring knowledge \cite{Cartaxo2016}.

Effectively transferring research findings into practice is a key aspect for bridging the gap between theory and practice. Budgen et al. \cite{budgen2020support} emphasized that SLRs alone do not effectively support the translation of research into practice, reinforcing the need for a more effective medium that simplifies research findings for use in the real world and industry. 

To achieve this, an evidence briefing should include the following sections, as suggested in \cite{Cartaxo2016,budgen2020support}:

\begin{itemize}
    \item \textbf{Title and Description}: The title guides the reader to the main focus of the briefing while maintaining simplicity to make it more appealing to practitioners. To enhance accessibility, technical terms that refer to research methods should be avoided, such as systematic review. Then, include a brief paragraph summarizing the briefing's purpose, providing relevant context for the reader.
    \item \textbf{Main Findings}: This section presents a concise summary of the key findings from the research paper, highlighting the most important insights.
    \item \textbf{Audience and References}: This section specifies the target audience and provides references to the original research for further reading.
\end{itemize}


Cartaxo \textit{et al.} \cite{Cartaxo2016} evaluated evidence briefings through surveys with StackExchange practitioners and systematic review authors. Practitioners perceived the briefings as clear and relevant, but noted limitations in addressing highly specific questions, \textit{i.e.}, they might miss details. SLR authors recognized their accessibility and effectiveness in summarizing key findings. Despite their benefits, their adoption remains limited. Automating their generation through LLMs may represent an opportunity to enhance scalability while maintaining their effectiveness as a means of knowledge transfer.



\subsection{LLMs for Research Synthesis}


Research synthesis has mostly relied on manual effort, with experts selecting, analyzing, and summarizing findings from multiple primary studies. As text summarization methods evolve, research synthesis becomes increasingly feasible, reducing the effort required to extract insights from the literature. Early summarization techniques employed extractive methods, selecting key sentences based on statistical features such as term frequency and sentence position. While these methods provided partial automation, researchers still needed to refine and interpret the extracted summaries manually \cite{zhang2024}. 

The advent of ML-powered models marked a shift toward more natural and concise summaries, capable of rephrasing content while preserving its meaning. Recent advancements in LLMs have further refined this process, leveraging deep contextual understanding and pre-trained knowledge to synthesize complex research findings \cite{zhang2024}. LLMs have proven to be a valuable tool in the Natural Language Processing field, enabling a wide range of tasks, including classification, text generation, question answering, sentiment analysis, and summarization.

Rosenfeld \textit{et al.}~\cite{rosenfeld2024building} introduced an AI-driven framework to summarize evidence to effectively communicate complex research findings to policymakers. Zhang et al.~\cite{zhang2024closing} compared different LLMs for evidence summarization, discussing how these models can be optimized to generate reliable, industry-relevant research summaries. Tang \textit{et al.}~\cite{tang2023evaluating} focused on benchmarking LLMs in summarizing research findings, making them directly applicable for generating evidence summaries for professionals in academia and industry. 

Despite these advancements, LLMs still present challenges in research synthesis. According to Tang \textit{et al.}~\cite{tang2023evaluating}, these include the need for effective prompt engineering, risks of generating hallucinated insights, and difficulty in capturing contextual nuances. They also highlight concerns regarding inconsistencies in LLM-generated summaries, warning that these models can sometimes produce overly convincing or uncertain statements, leading to potential misinformation.

\section{Experimental Study Plan}
\label{sec:experiments}

\begin{figure}[htp]
    \centering
    \includegraphics[width=0.48\textwidth]{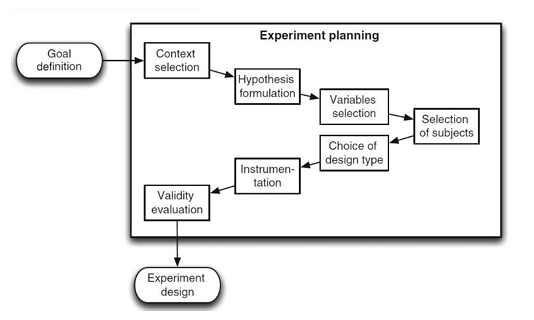}
    \caption{Experiment planning steps~\cite{wohlin2024experimentation}.}
    \label{fig:experimental-Study-plan}
\end{figure}

In order to determine whether LLM-generated evidence briefings can be effective within the SE domain to communicate research findings to practitioners, we designed our experimental investigation following the recommendations by Wohlin \textit{et al.}~\cite{wohlin2024experimentation}, see Figure \ref{fig:experimental-Study-plan}.

\subsection{Goal Definition}
\label{AA}

\edit{The main goal is to investigate whether automatically generated evidence briefings can be suitable for broader adoption. This goal was tailored into two specific sub-goals, defined using the Goal-Question-Metric (GQM) goal definition template proposed by Basili \textit{et al.}~\cite{Basili88} as follows.}

\edit{\textbf{GOAL 1}: \textit{Analyze} LLM-based evidence briefing generation \textit{for the purpose of} characterizing \textit{with} \textit{respect to} their content fidelity \textit{from the point of view} of researchers \textit{in the context of} evaluating automatically generated and human-made evidence briefings following the same structured template.}

\edit{\textbf{GOAL 2}: \textit{Analyze} LLM-based evidence briefing generation \textit{for the purpose of} characterizing \textit{with} \textit{respect to} their ease of understanding and perceived usefulness \textit{from the point of view} of practitioners \textit{in the context of} evaluating automatically generated and human-made evidence briefings following the same structured template.}

\edit{From GOAL 1, we derived the following research question (RQ):}
\begin{itemize}
    \item \textbf{RQ1}: How do researchers perceive the content fidelity of automatically generated and human-made evidence briefings?
\end{itemize}
\edit{From GOAL 2, we derived the following RQs:}
\begin{itemize}
    \item \textbf{RQ2}: How do practitioners perceive the ease of understanding of automatically generated and human-made evidence briefings?
    \item \textbf{RQ3}: How do practitioners perceive the usefulness of automatically generated and human-made evidence briefings?
\end{itemize}

\subsection{Context Selection}

The selected context consists of two independent offline controlled experiments, one involving researchers and the other industry practitioners, each designed to evaluate different aspects of automatically generated and human-made evidence briefings.

\begin{itemize}
     \item \textbf{Academic Experiment: Evaluating Content Fidelity}.
        In the first experiment, researchers will assess whether evidence briefings accurately capture the key information from the original research articles. Each participant will be presented with two examples of full-length research papers. For each research paper, the researcher will receive a randomly assigned version of the respective evidence briefing, either automatically generated or human-made. Researchers will then evaluate how well the briefing preserves the core findings and conclusions of the original work.
    \item \textbf{Industry Experiment: Evaluating Ease of Understanding and Usefulness}.
        The second experiment targets SE practitioners, focusing on how well they can understand and interpret the information presented in evidence briefings. Unlike researchers, practitioners will not be exposed to the full research articles. Instead, they will be randomly assigned to read one human-made and one automatically generated evidence briefing. Afterward, they will answer a set of questions designed to assess their perception of the briefing's ease of understanding and usefulness.
\end{itemize}

This division of evaluation roles reflects the expertise and perspectives of each group. Researchers are best suited to assess research content fidelity, as they are familiar with academic standards and able to verify how accurately the briefing represents the source material. Practitioners, on the other hand, are the intended consumers of these briefings, making them ideal evaluators for aspects such as ease of understanding and usefulness in their practice.

\subsection{Hypothesis Formulation}

$H0_1$: There is no difference between the researchers' perception of content fidelity for human-made and automatically generated evidence briefings.

$H0_2$: There is no difference between the practitioners' perception of ease of understanding for human-made and automatically generated evidence briefings.

$H0_3$:  There is no difference between the practitioners' perception of usefulness for human-made and automatically generated evidence briefings.

\subsection{Variable Selection}

The independent variable of interest (experimental factor) corresponds to the \textbf{type of evidence briefing} presented to the participants. \edit{Hence, the generation method is central to the study design}.
\begin{itemize}
     \item \textbf{Automatically generated evidence briefing}: Produced by an LLM-based system.
     \item \textbf{Human-made evidence briefing}: Manually produced by researchers.
\end{itemize}

Additional independent variables collected within the participant characterization include:  
\begin{itemize}
     \item \textbf{For researchers}: Familiarity with systematic reviews, academic role, and academic degree.
     \item \textbf{For practitioners}: Years of professional experience, role, academic degree, familiarity with each topic related to the selected evidence briefings.
\end{itemize}

The study defines three dependent variables, each corresponding to a specific research question:

\begin{itemize}
    \item \textbf{Content Fidelity}: Researchers' perception of how faithfully the evidence briefing represents the content of the original research article.
    \item \textbf{Ease of Understanding}: Practitioners’ assessment of the clarity and comprehensibility of the evidence briefing.
    \item \textbf{Usefulness}: Practitioners' perception of how useful the evidence briefing is for their professional practice.
\end{itemize}

Following recommendations for construct validity~\cite{creswell2017research}\cite{Kalinowski2024}, we measure each dependent variable using multiple items that reflect its theoretical components. For researchers, the dependent variable \textbf{content fidelity} is assessed through multiple items adapted from the error taxonomy proposed by Tang \textit{et al.}\cite{tang2023evaluating}. While Tang identified inconsistency types through qualitative coding, we incorporated these categories directly into the survey to enable a more focused evaluation. The items target three key dimensions: contradiction (alignment of conclusions), certainty illusion (accuracy of conveyed confidence), and fabricated content (presence of unsupported claims). This structured approach allows for a more detailed assessment of factual consistency in LLM-generated briefings. For practitioners, \textbf{ease of understanding} is evaluated with questions relating to the clarity of the language used, the logical structure of the text, and conciseness. This decomposition is further aligned with Creswell's~\cite{creswell2017research} discussion of readability, which emphasizes consistent terminology, logical flow between ideas, and coherence across sections. Finally, for \textbf{usefulness} in the context of SE, we assessed it based on the perceived relevance of the insights and recommendations for challenges or tasks related to professional practice and actionability (\textit{i.e.}, applicability in the work context).

The participant is presented with affirmative statements \edit{related to each of the items that compose the dependent variables} and asked to rate their level of agreement according to a 7-points Likert scale: 1 - Strongly Disagree, 2 - Disagree 3- Slightly Disagree, 4 - I neither agree nor disagree, 5 - Slightly Agree, 6 - Agree, and 7 - Strongly Agree.

Additionally, the evaluation questions include an optional open-text field, allowing participants to elaborate on their answers and enabling a complementary qualitative analysis. This follows a concurrent embedded mixed-methods strategy \cite{creswell2017research}, where qualitative insights help interpret and contextualize quantitative trends; we will make a composite assessment from the open questions to reside side by side with quantitative analysis and add depth and strengthen the overall validity of the findings.


\subsection{Selection of Subjects}
\label{SCM}
The SE researchers and practitioners \edit{for the two studies} will be selected based on convenience sampling and open invitations on social media. \edit{To answer RQ1 (content fidelity), in the first study, we will recruit a sample of SE researchers. To answer RQ2/RQ3 (understandability \& usefulness), in the second study, we will recruit a separate sample of industry practitioners}. They will be randomly assigned to the treatments, and we will use their independent variable characterization to apply blocking principles during the analysis. \edit{Furthermore, we will describe the samples in detail based on their characterization and carefully discuss their representativeness.} 

\edit{Despite using convenience sampling, it is noteworthy that participants are randomly assigned to the treatments and do not know which EB was generated by humans and which one was generated automatically. \textit{I.e.}, they could not intentionally favor a specific treatment, even if they would like to.}

\subsection{Choice of Design Type}

The goal is to investigate whether LLM-generated evidence briefings can effectively communicate research findings to practitioners. Therefore, we conduct a controlled experiment with researchers to evaluate the content fidelity of the generated evidence briefings, and another controlled experiment with practitioners to assess the perceived ease of understanding and usefulness. 

\begin{figure}[htp]
    \centering
    \includegraphics[width=0.48\textwidth]{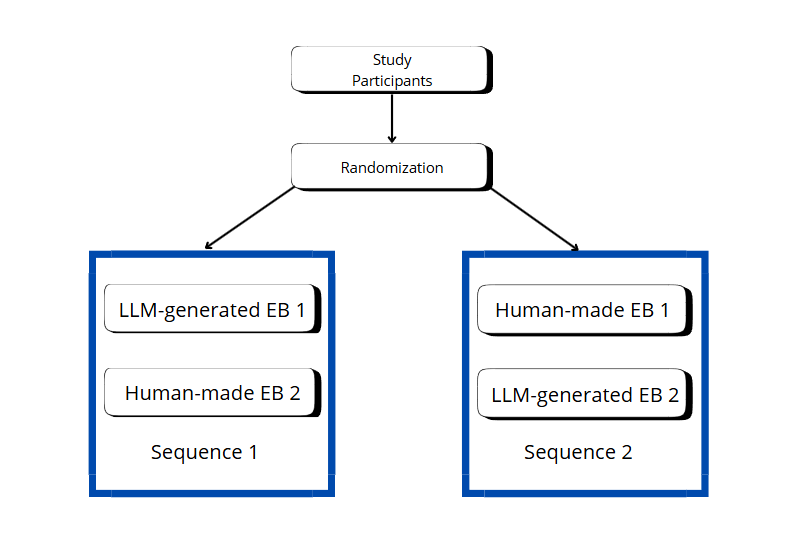}
    \caption{Crossover design}
    \label{fig:crossover design}
\end{figure}

We adopted a completely randomized one-factor, two-treatment crossover design for each of these experiments, as illustrated in Figure \ref{fig:crossover design}. This design was chosen to isolate inter-subject variability and learning effects. The factor concerns the type of evidence briefing and the treatments involved using LLM or manually generated ones. Furthermore, to avoid skewing the results, participants are randomly assigned to a pair of evidence briefings (LLM first and manual second, or manual first and LLM second). It is noteworthy that each of the two evidence briefings has both versions and that the human-made evidence briefings were generated in previous and independent research efforts~\cite{Cartaxo2016}. 

The experimental tasks for researchers concerned reading a research paper and the evidence briefing and evaluating content fidelity. For practitioners, the task concerned reading the evidence briefing and answering questions regarding ease of understanding and usefulness. \edit{In both cases, participants are not aware of the source of the briefing.}

\subsection{Instrumentation}

We designed separate online questionnaires using Google Forms for the researchers and for the practitioners. \edit{The study protocol and its instruments were officially approved by the CSU institutional ethics committee}. Each questionnaire consists of three sections: a consent form, a participant characterization form, and a main section with substantive questions that focus on the dependent variables. \edit{We evaluated both versions of each questionnaire \textit{(cf.} Figure \ref{fig:crossover design}) in a pilot study with two researchers and two practitioners, and no further adjustments were deemed necessary.}
\subsubsection{\textit{Consent form}} provides a general overview of the research and its objectives and reports ethical considerations as suggested by Badampudi \textit{et al.} ~\cite{eInformatica2022Art09}. The only anticipated risk concerns practitioners potentially being presented with information that is wrong or hallucinated.

\subsubsection{\textit{Participant Characterization Form}} This form was designed to collect the aforementioned demographic characterization information. It ensures that participants are properly categorized based on their research or industry experience, allowing for a more precise analysis of their responses.

\subsubsection{\textit{Main Section}} The main section of the questionnaire is designed to directly address the study's research questions, described in Section \ref{AA}. Participants are asked to evaluate evidence briefings related to two studies, one on pair programming~\cite{hannay2009pair} and one on definition of done~\cite{perkusich2017}. These briefings were produced manually in the context of the evidence briefings' usefulness evaluation~\cite{Cartaxo2018}. The studies were selected from a set of 56 evidence briefings because they summarize secondary studies \edit{with rigorous quality assessment procedures that were published in high-quality peer-reviewed venues}, on topics that are relevant to both academia and industry. As highlighted by Kuhrmann \textit{et al.}~\cite{kuhrmann2022agile}, agile software development is widely used by SE teams and is primarily shaped by its practices, reinforcing the importance of understanding the evidence related to such practices. \edit{While all participants evaluate briefings covering both topics  (pair programming and definition of done)}, researchers are additionally provided with the original research articles to assess content fidelity, whereas practitioners rely solely on the briefings to evaluate ease of understanding and perceived usefulness.

In order to automatically generate the evidence briefings, a tool was developed that leverages LLMs via API calls. We describe how the LLM was used according to the evaluation guidelines for empirical studies in SE involving LLMs~\cite{wagner2025evaluationguidelines}\footnote{https://llm-guidelines.org/guidelines/}, declaring the LLM usage and role, the model version and date, prompts and their development, and the tool architecture and supplemental data (see Table~\ref{tab:llm_config}) .

To ensure that the model produces structured and readable evidence briefings aligned with practitioner needs, we crafted a prompt using instruction-based prompting techniques, as outlined in prior work on prompt engineering~\cite{geroimenko2025key}. The prompt explicitly defines the task, introduces the evidence briefing format, and includes structured constraints to guide the LLM’s output. This approach aligns with key principles such as clarity, context provision, and output formatting, as described by Geroimenko~\cite{geroimenko2025key}.

To enhance the quality and coherence of the generated summaries, a Retrieval-Augmented Generation (RAG) structure was employed. This strategy aligns with the principle of providing contextual information as outlined by Geroimenko~\cite{geroimenko2025key}, which emphasizes that supplying relevant background helps guide the model towards more accurate, nuanced, and audience-appropriate responses. In our case, a ChromaDB database containing 54 human-made evidence briefings \edit{(the set of 56, excluding the two studies chosen for the experiment)} is used as the retrieval source. Before generating a briefing, snippets corresponding to each required section of the document are retrieved, offering the model grounded examples to adapt its tone, content structure, and language clarity. The tools' user interface also allows for the optional provision of keywords such as `Systematic Literature Review' or `Scrum' to improve the relevance of the retrieved context to the work to be summarized. We did not use this feature for the evidence briefings of our experiment.

\begin{table}[ht]
\renewcommand{\arraystretch}{1.2} 
\centering
\caption{LLM Configuration and Usage Summary}
\label{tab:llm_config}
\begin{tabular}{|l|p{5cm}|}
\hline
\textbf{Aspect} & \textbf{Details} \\
\hline
Model & GPT-4-o-mini (gpt-4-0125-preview) \\
Provider &  OpenAI API  \\
Access Date & March 2025 \\
Temperature & 0.5  \\
Top-p & 1.0  \\
Max Tokens & 1024  (output limit)\\
Prompting Strategy & Instruction-based prompting \\
RAG Mechanism & Retrieval via ChromaDB \\
Retrieved Examples & 54 human-made evidence briefings \\

\hline
\end{tabular}
\end{table}

To support further investigation and facilitate reproducibility, all materials used in our study are available in our immutable open science repository\footnote{\url{https://doi.org/10.5281/zenodo.15233046}}. These include the final prompt used, all the code and instructions necessary to run the tool, the human-made and LLM-generated Evidence Briefings, as well as the survey instruments.




\subsection{Threats to validity}

We discuss the threats to validity based on the classification by Wohlin et al.~\cite{wohlin2024experimentation}: \textit{construct}, \textit{internal}, \textit{external}, and \textit{conclusion} validity. For each category, we describe the key threats and our mitigation strategies.

\subsubsection*{Construct Validity}

A potential threat lies in the configuration of the LLM. Changes in the prompt formulation, temperature, top-p parameters, or retrieval corpus may affect the generated output and, consequently, the consistency of results. We mitigated this threat by following reporting guidelines for empirical studies in SE involving LLMs~\cite{wagner2025evaluationguidelines} and fully fixing and documenting all generation parameters, including prompt, temperature, top-p, and retrieval configuration. All outputs were archived and made available through our open science repository. 

Another concern relates to the adequacy of our survey instruments in capturing the theoretical constructs of interest. To address this, each construct was evaluated using multiple items to enhance measurement reliability. We adapted established evaluation taxonomies—such as Tang et al.~\cite{tang2023evaluating} for content fidelity—and followed guidelines for assessing readability and usefulness in software engineering contexts~\cite{creswell2017research}\cite{garousi2020}\cite{Kalinowski2024}. 

\subsubsection*{Internal Validity}

One internal threat is related to \textit{order effects} in the crossover design. Participants may be influenced by the sequence in which they receive the briefings. To mitigate this, we adopted a counterbalanced crossover design in which participants were randomly assigned to receive the human-made or LLM-generated briefing first. 

A second concern is \textit{carryover effects}, where exposure to the first briefing may influence the evaluation of the second. We minimized this threat by ensuring that each participant assessed two different topics (pair programming and definition of done), reducing priming effects between treatments.

\subsubsection*{External Validity}

The use of only two research papers may limit the generalizability of our findings to other areas of SE. To mitigate this, we selected papers that are both methodologically well-grounded and highly relevant to practitioners. Still, we acknowledge the need for replication across broader domains. 

Furthermore, differences in validation rigor between the human-made briefings represent an additional threat~\cite{Cartaxo2018}. The pair programming briefing had been previously validated~\cite{Cartaxo2016}, whereas the definition of done briefing was produced separately and did not undergo the same validation process. We acknowledge this as a limitation and interpret comparisons accordingly. 

Finally, a selection-treatment interaction may occur if participants lack familiarity with the topic of the briefing, leading to biased usefulness ratings. We collected self-reported familiarity data and will use it as a blocking variable during the analysis phase.

\subsubsection*{Conclusion Validity}

A threat to conclusion validity concerns statistical power. If the sample size is insufficient, the experiment may fail to detect significant effects. We will carefully consider sample size and representativeness throughout the experiment.

\section{Next Steps}
\label{sec:nextsteps}

The next steps of this research include conducting the experiment as specified in the registered report protocol, analyzing the data, testing the hypotheses using statistically appropriate methods, answering the research questions, discussing the results, and writing the complete journal paper to disseminate the findings to the community.


\bibliographystyle{plain}
\bibliography{refs}
\end{document}